# High Harmonic Generation from a Noble Metal

S. Gholam-Mirzaei[1,*], A. Korobenko[1], N. Haram[1], D. N. Purschke[1], S. Saha[2], A. Yu. Naumov[1], G. Vampa[1], D. M. Villeneuve[1], R. E. F. Silva[3], A. Staudte[1], A. Boltasseva[2], V. M. Shalaev[2], A. Jiménez-Galán[3,*] and P. B. Corkum[1,*]

[1]Joint Attosecond Science Laboratory (JASLab), National Research Council of Canada and University of Ottawa, Ottawa, ON, Canada
[2]Purdue University, School of Electrical & Computer Engineering and Birck Nanotechnology Center, West Lafayette, IN, USA
[3]Instituto de Ciencia de Materiales de Madrid, CSIC, Madrid, Spain
[*]Email: sgholamm@uottawa.ca, alvaro.jimenez@csic.es, pcorkum@uottawa.ca

**High-harmonic generation (HHG) in solids has typically been explored in transparent dielectrics and semiconductors. Metals have long been dismissed due to their strong reflectivity at infrared wavelengths. Here, we demonstrate HHG from silver—a noble metal—using few-cycle near-infrared laser pulses at near-normal incidence. Our results show that sub-cycle electron dynamics within the material's penetration depth can drive high-order harmonics, challenging the prevailing notion that metals are unsuited for infrared-driven strong-field processes.**

Despite silver's high reflectivity and large free-electron density, we observe nonperturbative harmonics extending into the extreme ultraviolet (up to 20 eV). Moreover, silver's multi-shot damage threshold (~30 TW/cm²) proves surprisingly high—comparable to large-bandgap dielectrics like magnesium oxide—thereby enabling intense strong-field processes in a metallic environment. Measuring the orientation dependence of the emitted harmonics reveals that the process arises from coherent electron dynamics in the crystal lattice, rather than from a plasma-driven mechanism. Time-dependent density-matrix simulations based on maximally localized Wannier functions show that lower-order harmonics predominantly originate from conduction electrons near the Fermi surface (s- and p-type orbitals), whereas higher harmonics rely on bound d-electron excitations. These findings establish metals—long thought unfavorable for HHG—as a promising platform for ultrafast strong-field physics, extending high-harmonic spectroscopy to regimes in which lattice order and plasma formation directly intersect. This work expands the frontier of solid-state HHG to all-metallic attosecond pulse generation and underscores the potential of metals as robust XUV sources for advanced attosecond metrology.

**Background**

When intense laser light interacts with condensed matter, it induces time-dependent electric currents in the material. These currents can emit coherent radiation at integer multiples of the laser frequency, extending well into the extreme ultraviolet (XUV) regime through high-harmonic generation (HHG) (*1*). HHG is not only a source of XUV light (*2*), it also reveals how materials responds to strong visible or infrared laser pulses on attosecond timescales (*3*). From this perspective, we might expect to measure high harmonic emission from all transparent solids near optical damage (*4*).

High harmonics were first observed in metal plasmas (*5*), where they arise from the field-driven oscillations of the free electron density at the plasma-vacuum interface, which acts as an oscillating mirror (*6, 7*). The generation of plasma harmonics is typically induced by laser fields with large incidence angles. Crucially, the process is insensitive to the crystal lattice or band structure, as these harmonics result from the collective electron behavior in a largely structureless plasma without memory of the condensed phase of the material (*8*).

A few years after this initial discovery, high harmonic generation (HHG) was demonstrated in gases (*2*). In this case, the harmonics were produced through recollision — a mechanism fundamentally linked to the coherent motion of electrons as they ionize and then recombine (*3, 9*). This direct link enabled probing electron motion in atoms and molecules on intrinsic attosecond timescales (*9*).

Twenty years later, HHG was extended to transparent solids (*10*). In these materials, two distinct emission mechanisms were identified: an intra-band mechanism (*11-14*), where the harmonics arise due to the motion of electrons within non-parabolic energy bands (*11, 12, 15*), and an inter-band mechanism (*13, 14, 16*), which is associated with the coherence between electrons and holes (*13, 14*), resembling the recollision harmonics in gases (*16*). Both mechanisms are highly sensitive to the coherent motion of carriers within the electronic bands (*13*). Consequently, they are also influenced by the crystal lattice and band structure of the material (*10, 17, 18*). This sensitivity to lattice and band structure stands in stark contrast to the harmonics generated from metal plasmas, where such microscopic details do not govern the emission mechanism.

HHG has since been demonstrated in a wide range of condensed matter systems (*1*), including dielectrics (*19-22*), semiconductors (*15, 17*), and highly correlated materials (*10, 23, 24*). These experiments have highlighted the versatility and potential of HHG for probing complex electronic structure (*13, 20, 25, 26*), coherent dynamics (*27*) and symmetries (*24, 28-30*) in solids.

Strikingly, HHG originating from the coherent electron motion within a noble metal's surface has not been reported so far. Such observation requires that the laser polarization mainly lies in the surface plane without significant plasma formation or surface ablation. Recently, by focusing intense few-cycle near-infrared (NIR) pulses near normal incidence, up to the 7$^{th}$ harmonic was generated from the surface of a thin film of titanium nitride (TiN) (*31*), a transition metal nitride (n-type, $\Delta E=3.4$ eV). The emission occurs for intensities near the optical damage threshold (10-12 TW/cm$^2$). The observations were consistent with an intra-band generation mechanism, underscoring the role of the Fermi-level electrons[10]. TiN exhibits several characteristics of metals due to local minima at the high symmetry points in the band structure. However, the lack of global minima in the band limits the band filling to only a small region of the conduction band minimum, leading to lower carrier concentration, reduced free-electron behavior and lower electron mobility compared to noble metals. Therefore, the strong field physics in TiN cannot be simply generalized to classical metals.

Here, we generate high harmonics from the quintessential metal. We use both epitaxial thin film and bulk single crystal Ag, and we observe a harmonic cut-off extending to the XUV. We find that the multi-shot (~10$^5$-shots) damage threshold for silver exceeds that of TiN and is comparable to magnesium oxide – a transparent solid often selected in HHG studies due to its high damage threshold and a promising candidate for producing intense XUV.

Unlike recollision harmonics from transparent solids, high-harmonic emission from a metal remains efficient even when driven by circularly polarized light. This observation also reveals that the damage threshold is determined by the peak field strength rather than the average power.

We observe that the high harmonic yield strongly depends on the crystallographic orientation, indicating that the electronic states of the metal are responsible for the emission. This result contrasts with expectations for Ag, characterized by a high electron density (n=5.86×10$^{28}$ m$^{-3}$) (*32*) and high reflectivity (99.5% at 800 nm) (*33, 34*).

Using time-dependent reduced density matrix calculations (*35*) in Ag, we show that low-order harmonic emission is mainly governed by the motion of Fermi level electrons., High-order harmonics, on the other hand, originate instead from coherences between low-lying d-type electronic bands and high-lying p-type electronic bands, proving the importance of bound electrons in the generation of high harmonics from metals.

**HHG Driven by Intense NIR Few-Cycle Pulses:**

In the experiment, sub-two-cycle NIR pulses are obtained by compressing 30 fs, 1 mJ pulses ($\lambda_{center}$=780 nm) of a commercial Ti:Sapphire laser amplifier (Femtopower) using a neon-filled hollow-core fiber and chirped mirror compressor. As illustrated in Figure 1a), near-normal incidence few-cycle pulses are focused onto the 65 nm-thick Ag that was epitaxially grown on a silicon substrate (see Methods). The reflected light containing high harmonic emission passes through a 300 μm-wide slit, and the harmonics are imaged in the far-field using an XUV grating and an imaging multi-channel plate (MCP). We control the intensity applied to the sample by using a broadband half waveplate and a broadband wire grid polarizer.

Figure 1b) shows the harmonics spectra as a function of driving field intensity. At intensities around 2.9×10$^{13}$ TW/cm$^2$, we observe changes in the spatial profile of the harmonics, indicating the onset of multi-shot damage. For intensities above 3×10$^{13}$ W/cm$^2$, the harmonic spectral intensity decreases drastically and eventually vanishes (yellow line in Fig.1b). Upon re-exposing the same spot on the sample, we were unable to generate the same harmonic yield. This indicates an irreversible damage mechanism. Hence, we take this value as the damage threshold of Ag, which is comparable to that of magnesium oxide exposed to NIR few-cycle pulses and surpasses the measured damage threshold in TiN (*31*) by more than a factor of two, illustrated by the dashed teal line in Figure 1b). We observe the high harmonic cut-off energy at ~20 eV (11$^{th}$ harmonic) before multi-shot damage occurs. We also observe a stronger high harmonic yield from thin film Ag than from thin film TiN for the same pulse duration and intensity (see Supplementary Material).

We relate this effect to the higher reflectivity and possibly higher electron density in Ag.

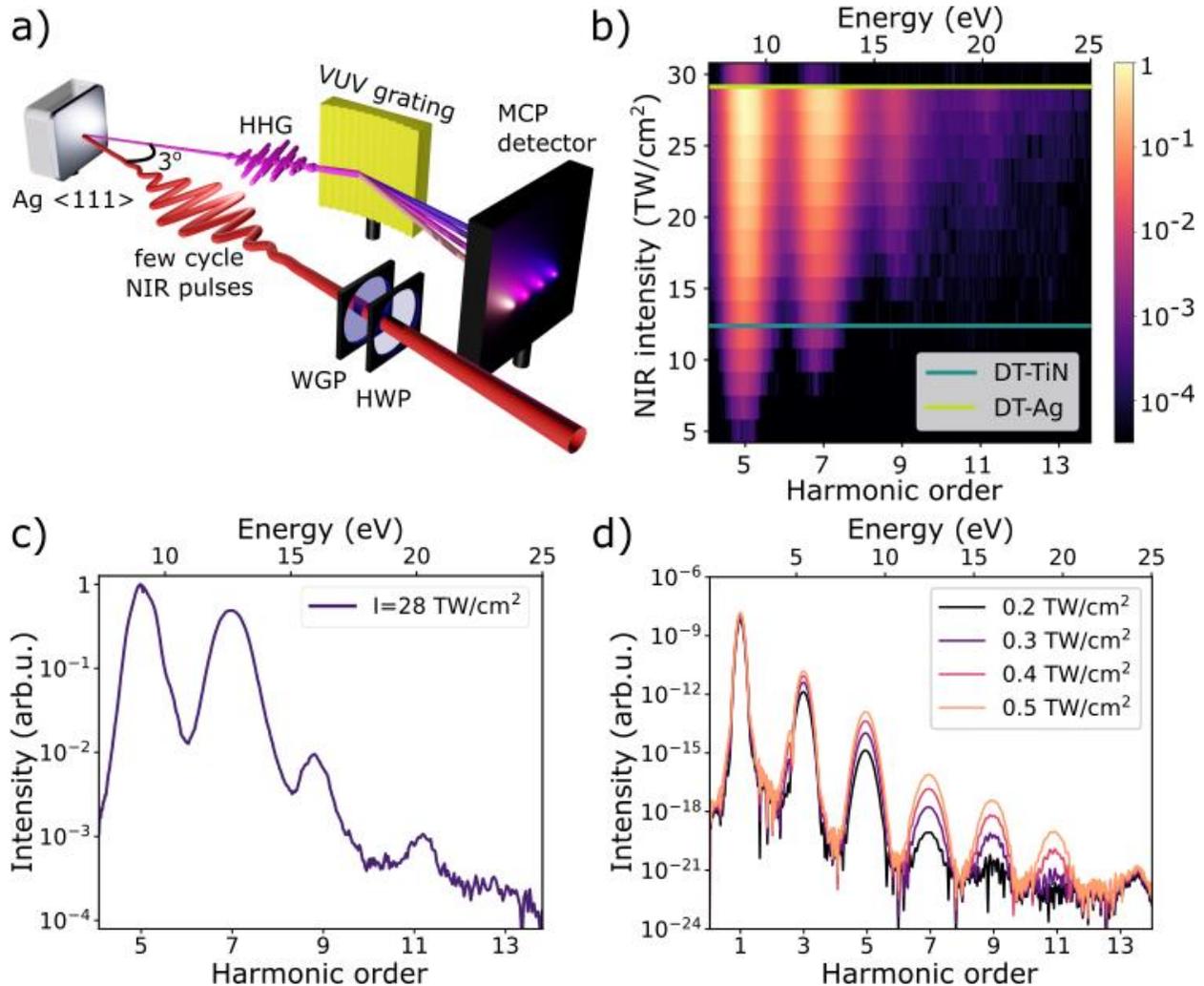

Fig. 1: a) Optical setup: a sub-two-cycle NIR pulse is focused on Ag target mounted on a tailored motorized stage. A broadband half wave plate (HWP) and a broadband wire grid polarizer (WGP) are used for preconditioning the driving beam's intensity. The emitted harmonics are imaged using a microchannel plate (MCP) detector and CCD camera. b) Harmonics spectral intensity as a function of incident beam's intensity. The solid yellow line defines the damage threshold intensity. For comparison, the damage threshold of TiN is shown by the dashed teal line. c) HHG spectrum at near optical damage intensity. The cut off energy is around 20 eV. d) Theoretically calculated HHG spectra corresponding to various peak intensities inside the metal (see Methods).

Figure 1c) presents the emitted harmonic spectrum and cut off at the intensity slightly below the damage threshold. When light is incident onto a highly reflective metal, the high density of electrons reflects most of the light. For our broadband NIR, it is reasonable to assume that 1% of the incident field is absorbed. If surface damage occurs at $I_0 \sim 3\times 10^{13}$ W/cm$^2$ in vacuum, then the maximum achievable intensity inside the metal is approximately $I_{int}$~$3\times10^{11}$ W/cm². Indeed, Figure 1d) depicts the theoretically calculated HHG spectra of Ag for different pulse intensity values (see Methods). At $I_{int}$~$3\times10^{11}$ W/cm², the theoretical calculation well describes the experimentally obtained spectral features and cut-off.

**Intensity Scaling of Harmonics**

We now study the scaling of each harmonic order as a function of the incident light intensity as illustrated for harmonics 5, 7, and 9 in Figure 2a). The solid lines correspond to the perturbative power law scaling $I_n = I_0^n$ where n stands for the order of nonlinearity. From the measurement, it is evident that harmonics deviate from the perturbative scaling above 18 TW/cm². This deviation is slight for harmonic 5 and 7 while it is more pronounced for harmonic 9. Figure 2b) shows the calculated intensity scaling of harmonics for an incident NIR pulse intensity of $I_{int}$ =0.3 TW/cm². Harmonic 7 shows a small deviation from the perturbative scaling while harmonic 9 shows a clear deviation, in line with the experimental findings.

**HHG Driven by Circularly Polarized Light**

Next, we use a broadband half wave plate (HWP) and a broadband quarter wave plate (QWP) to alter the polarization state of the incident NIR pulses from linear to circular polarization, such that the major axis of the polarization ellipse stays fixed to the optic axis of the crystal where we observe the maximum yield of the emitted harmonics. Figure 3c) shows the measured yield of harmonic 5 and harmonic 7 for linear (red curve) and circular (black curve) polarizations of the driver, at the same incident average power. The yield does not change significantly; an observation that is reproduced in our simulations (Fig. 2d). This is in stark contrast to standard dielectrics and semiconductors in which harmonics emission is quenched when driven by circularly polarized light. The suppression of harmonic 9 reflects the six-fold symmetry of the crystal lattice of Ag (111), as shown in Figure 3a)(*36*).

Interestingly, we find that the sample withstands higher intensities for circular polarization than for linear polarization. We can increase the incident light intensity by 25% compared to linearly polarized light (green curves in Figs. 2c, d), finding that the harmonics remain bright up to the new damage threshold. This observation suggests that multi-shot damage is determined by the instantaneous peak electric field strength, which is twice in the linear case, and not the cycle-averaged power, which is the same for both polarizations.

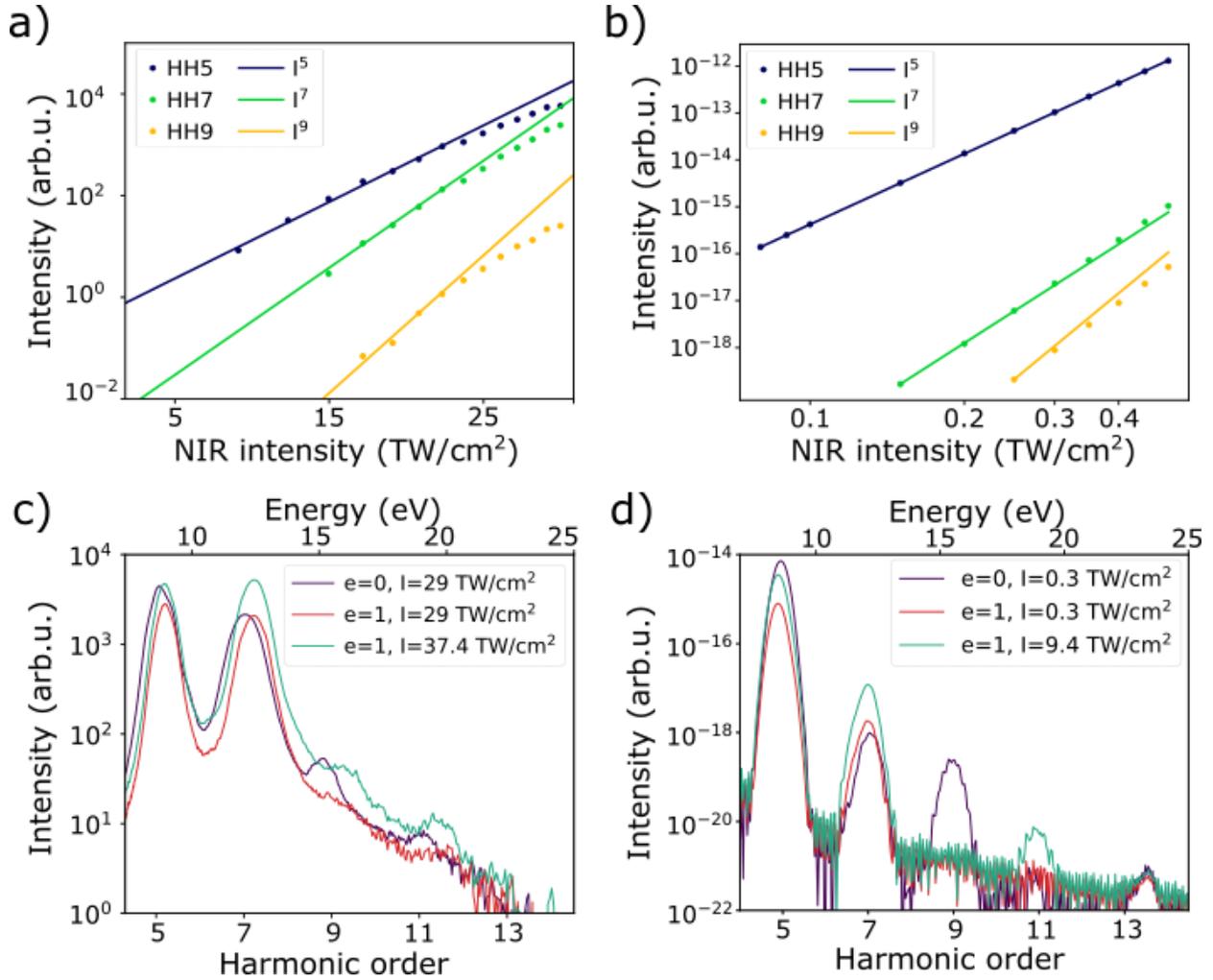

Fig. 2: a) Experimentally obtained intensity scaling of harmonics from Ag. The dots show the measured intensity of the harmonics as a function of peak intensity, while the solid lines show the power-law scaling. b) Calculated intensity scaling of harmonics. c) Measured harmonic spectrum for linearly (purple) and circularly (red) polarized driving pulses at an intensity slightly below the multi-shot damage intensity (I=29 TW/cm$^2$). The harmonic yield is enhanced in circularly polarized pulses (teal) beyond the damage threshold intensity in linear light. d) Calculated harmonic spectrum for NIR driving field with linear (purple), and circular (red) polarization at I = 0.3 TW/cm$^2$, estimated as the near-damage intensity value inside Ag. The teal line demonstrates further enhancement of the yield of harmonics for intensities exceeding the damage threshold (I = 0.4 TW/cm$^2$) when using circularly polarized light.

**High Harmonics Spectroscopy of Lattice Structure**

To confirm that the harmonic emission arises from electrons in the lattice potential of the Ag crystal, we measure the harmonic dependence on the lattice symmetry. For this purpose, we use an additional half-wave plate to rotate the polarization of the driving relative to [110] direction in Ag (111) plane. The geometry of the Ag (111) plane is depicted in Figure 3a). The relative angle between the laser polarization and [110] directions is defined by θ. At each θ, we record the harmonics spectra. As shown in Figure 3b), both harmonic 5 and harmonic 7 harmonics exhibit the six-fold symmetry of the (111) plane of the crystal. The intensity modulation of harmonics 7 is also more pronounced than for harmonic 5, which reflects interferences between a larger number of recombination channels in the Brillouin zone or, equivalently, a larger number of lattice orbital contributions to the total current (*24*). Figure 2c) displays the theoretical simulation, which reproduces the experimental observations, reaffirming that high-harmonic emission in silver stems from crystal-lattice electrons rather than plasma dynamics. confirming our interpretation that harmonics are sensitive to the lattice potential and are not generated from metal plasma.

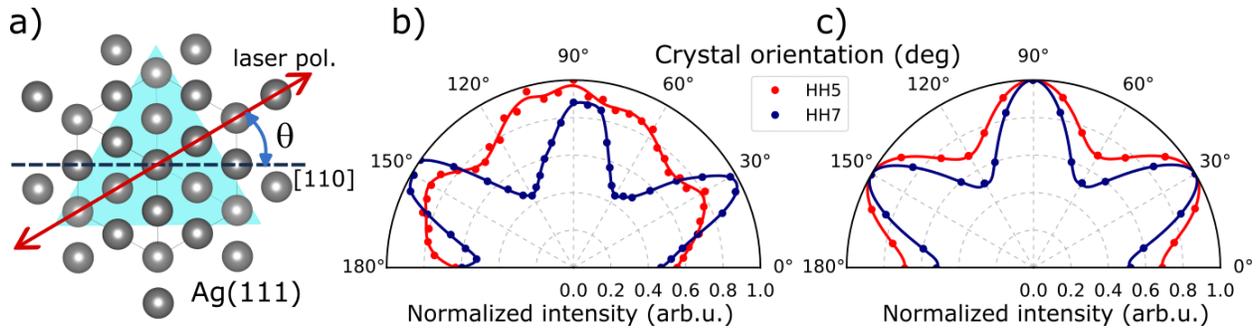

Fig. 3. a) orientation dependence of harmonic spectra from epi-Ag. a) Ag crystal structure in <111> plane. b) Experimental orientation-dependent spectra for 5$^{th}$ and 7$^{th}$ harmonics. c) theoretical orientation dependent spectra.

As discussed earlier, a metal's reflectivity is largely determined by its electron density and penetration depth is on the order of a few tens of nanometers in the NIR spectral domain. For a collisionless plasma, the penetration depth $\delta_e$ is defined by $\delta_e = \frac{c}{\text{Im}(n)\,\omega_e}$, where c is the light velocity and $\omega_e$ is the plasma frequency. With an electron density of $n_e$=5.86×10$^{28}$ m$^{-3}$ (*32*), the plasma frequency of Ag is about 8.9 eV-9.1 eV, corresponding to the plasma penetration depth of ~20-35 nm for a broadband NIR pulse (λ=600-900 nm). Therefore, our 65 nm thin film Ag can be effectively treated as a bulk medium and we expect that the silicon substrate interface has minimal influence on the pulse propagation.

In our multi-shot experiments, each laser pulse frees some of the bound electrons in a cumulative process, gradually increasing the absorption and heating the material (*37*). The Fermi energy of electrons in silver is 5.49 eV corresponding to a Fermi velocity of $v_f=1.4 \times 10^6$ m/s. If we consider the time between collisions, i.e., the scattering time, of $\tau \sim 20\text{-}40$ fs, the Drude–Sommerfeld semiclassical picture of electron transport in metals suggests that the average distance traveled between collisions, i.e., the mean free path, for a Fermi velocity electron is $l = v_f \tau \approx 30\text{-}60$ nm (*32*).

On the other hand, from the calculated Fermi velocity, we can find that the electron only travels about 3.5 nm for each cycle of 800 nm light, hence we expect total displacement of ~7-8 nm within the duration of each pulse. This means that the damage is caused by shot-by-shot build-up of absorption and the thermal response of the material (*37*).

Reducing the number of shots will allow the metal surface to withstand higher field strength, where the absorption and the penetrated field in the material will increase further. Ultimately at the single-shot damage threshold, the enhancement in the local field causes the transition from a crystalline metal state to a plasma-like state.

By tracking how the harmonic yield—and its dependence on polarization rotation—evolves as the reflectivity drops, one can probe the loss of crystalline structure and the emergence of plasma behavior. In other words, the orientation dependence of harmonics serves as a sensitive measure of ultrafast structural changes within the metal, allowing us to monitor how the lattice potential is "washed out" and replaced by a dense plasma as damage thresholds are approached.

**Role of Fermi level and Bound Electrons in the HHG Process**

In order to gain insight on the HHG mechanism, we turn to a time-dependent density matrix simulation which uses the ground state energies and dipole couplings of silver (*38, 39*)(see Methods). Our calculations include the 9 bands closest to the Fermi surface (see Fig.4a), corresponding to Wannier functions obtained from projections onto the 5*s*, 5*p* and 4*d* orbitals. Figures 4b-d) show the projected density of states. Lower valence bands are mostly composed of bound 4*d* electrons, the bands between 0 and 10 eV above the Fermi energy are contributed by both 5*s* and 5*p* electrons, and those higher than 10 eV above the Fermi energy are mostly 5*p* electrons.

Is HHG dynamics in metals driven only by conduction band electrons (i.e., *s* and *p* orbitals), or is there also a contribution of bound electrons (i.e., *d* orbitals)? Previous works on HHG in solids have focused on gaining physical insight of the harmonic generation process by separating the total current into an intraband and interband contribution. As we show in Supplementary Material, performing this separation here is not useful since interband currents can easily originate from conduction electrons driven towards the band crossings around 5eV above the Fermi energy (see Fig.4a). Instead, we calculate the energy-resolved electronic populations as a function of time in the presence of the strong laser field. The resulting population at the end of the interaction with the field is shown in Figure 4e) (see Supplementary Material for other times during the laser-crystal interaction). We see a clear depletion of bound electrons, indicating their participation in the dynamics, which goes in conjunction with the population of higher energies.

This observation is corroborated further by separating the total current into contributions of the different *s*, *p* and *d* orbitals (*35*). While intra- and interband currents are mixed due to the multiple band crossings, the localized Wannier orbitals are well separated into distinct energy regions (see Figs.4a-d). In Figures 4f,g), we show the high harmonic spectrum calculated by neglecting the contributions of *d* and *s* orbitals to the total current, respectively. While neglecting *d* orbitals has almost no effect on the lower harmonics (1, 3 and 5), its effect is large for the higher harmonics (7 and 9). Conversely, neglecting *s* orbitals has almost no effect on the high harmonics, but has a large effect for the lower harmonics. Hence, we find that lower order harmonics are dominated by dynamics of conduction electrons while high harmonics (7 and 9) are dominated by coherences from bound electrons.

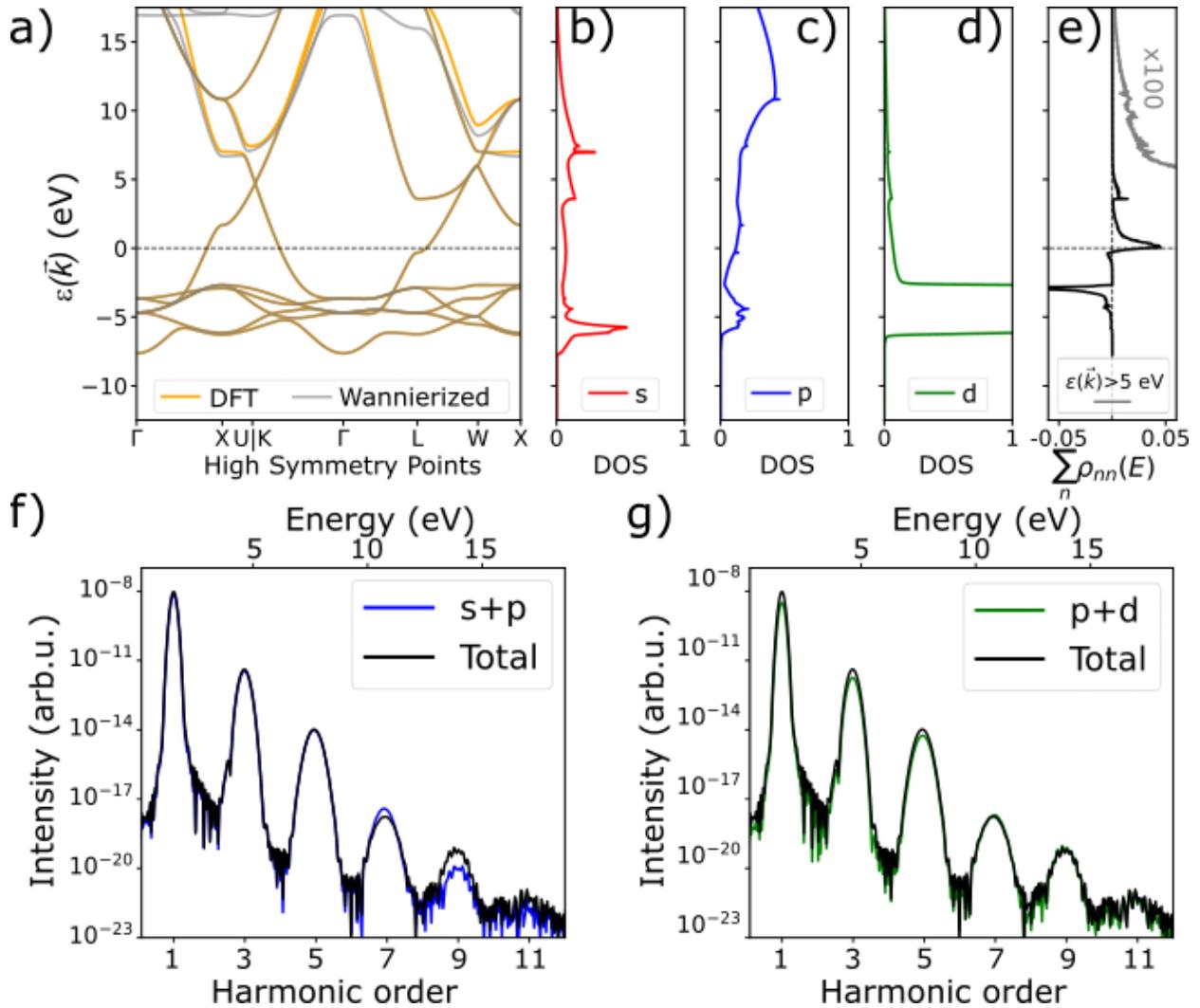

Fig. 4: (a) DFT calculated band structure (grey) and the Wannierized band structure (orange). The latter is calculated by projecting onto the *s*, *p* and *d* orbitals of Ag. The dashed line indicates the Fermi level. (b-d) Density of states projected onto the *s* (red), *p* (blue) and *d* (green) orbitals of Ag. (e) Energy-resolved population after the pulse interaction, relative to the initial population. Grey line shows the same for $\varepsilon(k) > 5$ eV but enhanced 100 times to show the higher energy population distribution. (f,g) Total HHG spectrum (black) and HHG spectrum calculated by neglecting the contribution of the current generated by *d* orbitals (panel f, blue) and *s* orbitals (panel g, green).

**Conclusion and Outlook**

We have reported observation of high harmonic emission from silver -- a noble metal. Our measurements unambiguously show that such harmonic emission is produced from electron dynamics in the periodic lattice, extending ultrafast strong-field spectroscopy and high harmonic generation to metals. Our findings are supported by ab-initio theory with which we have demonstrated the role of bound electrons in the generation of high harmonics from metals. Such bound electron excitation is possibly responsible for the observed high damage threshold in silver.

High harmonics from metals also opens new approaches to measuring the dynamics of material melting. During the time of our 5 fs pulse the atoms in the lattice will barely move. Angle-dependent emission is a measure of this order, giving us a tool to measure decoherence and loss of order in metals and the emergence of plasma harmonics as metals melt and the interface between vacuum and the material governs the high harmonic emission process. We expect to be able to distinguish the source of the harmonics by the angle dependence or the spectral phase of the harmonics.

We expect high-harmonic emission to be the universal response of all metals at high intensity, as in transparent solids (*10*), gases (*2*) and liquids (*40*), thus adding to a set of tools that is emerging for integrated XUV photonics -- tools such as structured solids for focusing (*38*) and XUV generation in the solid state (*41*).

**Methods**

**Sample Preparation**

Titanium nitride films were deposited on magnesium oxide substrates using DC magnetron sputtering at 800°C. A 99.995% pure titanium target of 2 inches diameter was used. The DC power was set at 200 W. To maintain high purity of the grown films, the chamber was pumped down to $3\times10^{-8}$ Torr before deposition and backfilled with Argon gas to 5 mTorr during the sputtering process. The throw length was 20 cm, ensuring a uniform thickness of the grown TiN layer throughout the 1 cm by 1 cm MgO substrate. After heating, the pressure increased to $1.2\times10^{-7}$ Torr. An argon-nitrogen mixture at a rate of 4 sccm/6 sccm was flowed into the chamber. The deposition rate was 2.2 Å min-1.

The epitaxial silver was grown on silicon. Ag thin films were deposited on prime-grade degenerately doped Si (111) using 10 kW e-beam evaporator (Angstrom Engineering) with a base pressure lower than $3 \times 10^{-8}$ Torr. The wafers were first cleaned in a 2:1 sulfuric acid: hydrogen peroxide solution (80 °C), followed by further cleaning in isopropanol to eliminate organics. Finally, the wafers were placed in 49% hydrofluoric acid for approximately 20 s to remove the native oxide layer. After oxide removal, the wafers were immediately transferred into the evaporation tool and the system was pumped down to limit native oxide growth. All films were grown using 5N (99.999%) pure silver. Films were deposited with rate of 0.5–10 Å·s$^{-1}$ measured with quartz monitor at approximate source to substrate distance of 30 cm. Deposition is done in two steps using SCULL process. The details of the growth process can be found in (*42*).

**Sample Characterization via XRD**

X-ray diffraction (XRD) measurements of the silver films on silicon were made with a Bruker D8 Advance X-ray diffractometer equipped with a sealed Cu tube source ($\lambda = 1.541836$ Å). Scans were performed with a 2θ range from 10 to 90° with a step size of 0.1° and a step time of 5 s. As shown in SI 2. the two major peaks at 29° and 38° correspond to epitaxial Ag (111) film and the lattice matched silicon (111) substrate respectively.

**Optical Setup**

In the experiment, a Ti:Sapphire laser amplifier (Femtolasers Femtopower Compact Pro Amplifier) operating at 1 kHz repetition rate, provides pulses with 30 fs temporal bandwidth and up to 1.5 mJ pulse energy. We take advantage of an argon-filled hollow-core fiber to spectrally broaden 300 μJ pulses and further compress them to ~5 fs FWHM duration using chirped-mirror compressor set up. Both uncompressed and compressed pulses are characterized using the dispersion scan technique .

A set of half waveplate and polarizers is used to control the intensity and the polarization of the driving laser beam. Next, by using a focusing mirror (500 mm FL) inside a vacuum chamber, the few-cycle pulses are focused on the metallic target with near-normal angle of incidence (~1.5°). The beam profile at the focal spot was characterized with a Thorlabs CMOS camera and found to have a beam waist diameter of 140 μm (FWHM). To obtain higher intensity, the beam waist size can be further reduced by altering the incident beam's size and the incoming energy. The peak intensity in the vacuum is calculated from the measured pulse energy, the pulse duration, and beam waist diameter.

In the detection chamber, the harmonics are dispersed by a 300 grooves/mm laminar-type replica diffraction grating (Shimadzu) and sent to a MCP imaged by a CMOS camera (PCO-flim).

The imaging MCP spectrometer is calibrated by exploiting XUV harmonics emission from an MgO surface using the uncompressed (25-30 fs) pulses.

The grating response for distinct polarization states are found by using MgO crystal (100) and is accounted for in the orientation-dependent harmonics yield (*18*).

**Band Structure Calculations and the Time-dependent HHG Simulations**

We performed numerical simulations of the experiment using the density matrix formalism. We computed the equilibrium structure using density functional theory(*43*) with the code Quantum Espresso (*39*). To avoid numerical problems during the propagation stemming from the gauge freedom of the phases of the Bloch vectors, we projected onto a basis of maximally-localized Wannier orbitals using the Wannier90 code (*35, 44*), obtaining the tight binding expression for the Hamiltonian and the dipole couplings of the system. The projection was made onto the *s*, *p* and *d* orbitals of Ag, giving a total of 9 bands, which correctly describes the system up to ~15 eV above the Fermi energy (see Supplementary Material Fig. 1). The density matrix of the 9-orbital Ag system was then propagated in the presence of the strong laser field to obtain the high harmonic response (*35*).

The propagation vector of the laser field was taken to be the same as in the experiment, i.e., along the (1,1,1) direction, with polarization in the (111) plane of the crystal. The laser field wavelength was chosen the same as in the experiment, $\lambda$=800 nm, with a Gaussian envelope of 5 fs of full-width at half maximum. The HHG spectrum was converged on a grid of 120x120x120 k-points and a time step of 0.4 atomic units. We confirmed the six-fold symmetry of the (111) crystal direction. The estimation of the laser intensity inside the material given in the main text was further corroborated by comparing the experimental and theoretical harmonic cut-off, as discussed in Figure 1 of the main text.

**Acknowledgement**

The authors acknowledge A. S. Baburin, I. A. Ryzhikov, and I. A. Rodionov's contribution from Bauman Moscow State Technical University, Moscow, who grew the epitaxial silver film. The authors are thankful to Tom Lacelle for carrying out the characterization of epitaxial Ag samples via XRD. We also thank D. Crane and R. Kroeker for technical support in the lab.

**Funding**

S.G.-M and P. B.C acknowledge support from Army Research Office through grant No. W911NF-24-1-0214, and the Joint (University of Ottawa and National Research Council of Canada) Center for Extreme Photonics (JCEP). A.J.-G. acknowledges support from the Talento Comunidad de Madrid Fellowship 2022-T1/IND-24102 and the Spanish Ministry of Science, Innovation and Universities through grant reference PID2023-146676NA-I00.

**Contributions**

P.B.C., S.G.-M., and A.J.-G. conceived the project. S.G.-M. conducted the measurements and analyzed the data. A.K., N.H., and G.V., D.M.V. and A.S. helped with the data acquisition. D.N.P. contributed to the data analysis.

S.S. prepared the epitaxial metallic sample. A.J.-G performed the theoretical analysis and time-dependent simulation. R.E.F.S. developed the time-dependent code. S.G.-M., A.J.-G., and P.B.C. wrote the manuscript. P.B.C. supervised the research. All authors discussed the obtained results and contributed to the manuscript.